\documentclass[10pt,preprint]{aastex}
\begin{document}

\title{And in the Darkness Bind Them: Equatorial Rings, B[e]
Supergiants, and the Waists of Bipolar Nebulae\altaffilmark{1}}

\author{Nathan Smith\altaffilmark{2,3,4,5}, John
Bally\altaffilmark{3,5}, and Josh Walawender\altaffilmark{3,5,6}}

\altaffiltext{1}{Based in part on observations made at the Clay
Telescope of the Magellan Observatory, a joint facility of the
Carnegie Observatories, Harvard University, the Massachusetts
Institute of Technology, the University of Arizona, and the University
of Michigan.}

\altaffiltext{2}{Astronomy Department, University of California, 601
Campbell Hall, Berkeley, CA 94720; nathans@astro.berkeley.edu}

\altaffiltext{3}{Center for Astrophysics and Space Astronomy, University of
Colorado, 389 UCB, Boulder, CO 80309}

\altaffiltext{4}{Visiting Astronomer at the New Technology Telescope
of the European Southern Observatory, La Silla, Chile.}

\altaffiltext{5}{Visiting Astronomer, Cerro Tololo Inter-American
Observatory, National Optical Astronomy Observatory, operated by the
Association of Universities for Research in Astronomy, Inc., under
cooperative agreement with the National Science Foundation.}

\altaffiltext{6}{Institute for Astronomy, 640 North A'ohoku Place,
Hilo, HI 96720-2700}

\begin{abstract}

We report the discovery of two new circumstellar ring nebulae in the
western Carina Nebula, and we discuss their significance in stellar
evolution.  The brighter of the two new objects, SBW1, resembles a
lidless staring eye and encircles a B1.5 Iab supergiant.  Although
seen in Carina, its luminosity class and radial velocity imply a
larger distance of $\sim$7 kpc in the far Carina arm.  At that
distance its size and shape are nearly identical to the equatorial
ring around SN1987A, but SBW1's low N abundance indicates that the
ring was excreted without its star passing through a red supergiant
phase.  The fainter object, SBW2, is a more distorted ring, is N-rich,
and is peculiar in that its central star seems to be invisible.  We
discuss the implications of these two new nebulae in context with
other circumstellar rings such as those around SN1987A, Sher~25,
HD~168625, RY~Scuti, WeBo1, SuWt2, and others.  The ring bearers fall
into two groups: Five rings surround hot supergiants, and it is
striking that all except for the one known binary are carbon copies of
the ring around SN1987A.  We propose a link between these B supergiant
rings and B[e] supergiants, where the large spatially-resolved rings
derive from the same material that would have given rise to emission
lines during the earlier B[e] phase, when it was much closer to the
star.  The remaining four rings surround evolved intermediate-mass
stars; all members of this ring fellowship are close binaries, hinting
that binary interactions govern the forging of such rings.  Two-thirds
of our sample are found in or near giant H~{\sc ii} regions.  We
estimate that there may be several thousand more dark rings in the
Galaxy, but we are scarcely aware of their existence -- either because
they are only illuminated in precious few circumstances or because of
selection effects.  For intermediate-mass stars, these rings might be
the pre-existing equatorial density enhancements invoked to bind the
waists of bipolar nebulae.

\end{abstract}


\section{INTRODUCTION}

Non-spherical mass loss is a key ingredient for understanding the
roles of rotation, magnetic fields, and close binary interactions in
various stages of stellar evolution.  Bipolar structure is common in
nebulae surrounding evolved stars across the HR Diagram, most notably
in planetary nebulae (Balick \& Frank 2002) and in more massive hot
stars like luminous blue variables (LBVs).  Many bipolar nebulae have
a dense torus or ring at their pinched waist where the polar lobes
meet at the equator, as seen in prototypical objects like
$\eta$~Carinae (Smith et al.\ 2002$a$) and in dramatic {\it Hubble
Space Telescope} ({\it HST}) images of planetary nebulae like MyCn 18
(Sahai et al.\ 1999), NGC2346,\footnote{See the Hubble Heritage image
at {\tt http://heritage.stsci.edu/1999/37/index.html}.} and NGC7027
(Latter et al.\ 2000).

Perhaps the most outstanding and familiar example of such equatorial
structure is the inner ring around SN1987A (Plait et al.\ 1995;
Burrows et al.\ 1995; Crotts \& Heathcote 1991; Crotts et al.\ 1989).
In SN1987A as well as planetary nebulae, the prevailing interpretation
for the formation of these equatorial rings and bipolar nebulae has
been that a fast low-density wind expands into an asymmetric slow
dense wind from a previous red giant or supergiant phase (e.g.,
Blondin \& Lundqvist 1993; Martin \& Arnett 1995; Collins et al.\
1999; Frank \& Mellema 1994; Frank 1999).  This requires that the slow
wind had a strong equatorial density enhancement (i.e. a disk) that
can be swept-up to form a ring, but the origin of this putative
pre-existing disk is unclear, and evidence of them around red
supergiants is non-existent.

The formation of equatorial disks by single stars has been of continued
interest to the Be star community, since these stars are rapid
rotators.  B[e] supergiants are suspected to be their high luminosity
analogs (e.g., Zickgraf et al.\ 1986).  The wind-compressed disk model
(Bjorkman \& Cassinelli 1993) suggests that rapidly rotating stars
will form disks as material ejected at mid-latitudes is directed
toward the equator.  While such disks are inhibited in hot stars with
line-driven winds because of the effects of gravity darkening and
velocity-dependent forces (Owocki et al.\ 1996; Cranmer \& Owocki
1995), wind compression may still operate in other
circumstances. Ignace et al.\ (1996) suggested that cool stars may
experience moderate compression, while Smith \& Townsend (2007)
propose that thin disks can form in continuum-driven LBV eruptions.
Changes in ionization with latitude on a gravity-darkened rotating
star has been suggested to play a role in disk formation as well
(e.g., Lamers \& Pauldrach 1991; Cure et al.\ 2005).  However, any
disk formation around cool stars requires rapid rotation much in
excess of rotation rates expected for the distended envelopes of red
giants and supergiants.  Indeed, Collins et al.\ (1999) estimated that
while in the red supergiant phase, the progenitor of SN1987A must have
been rotating at $\ga$0.3 of its critical breakup speed to produce the
observed nebular morphology --- which is impossible for normal
single-star evolution --- so they conclude that the red supergiant was
most likely spun-up by swallowing a companion.  Thus, close binary
interactions and mergers are often taken as potential sources of
angular momentum to focus material toward the equator.  Magnetic
fields may provide yet another avenue for the formation of equatorial
disks or rings (ud-Doula \& Owocki 2002; Washimi et al.\ 1996;
Townsend \& Owocki 2005; Matt \& Balick 2004).

The equatorial ring around SN1987A is quite unusual in that it really
is a hollow ring and not just a limb-brightened shell or pinched waist
of an hourglass.\footnote{Here we are using the word ``ring'' to
denote a thin toroid in 3-dimensional space, like a wedding ring, as
distinguished from an apparent ring on the sky caused by a
limb-brightened spherical shell like the ``ring nebulae'' around
Wolf-Rayet stars, for example.}  A virtual twin of SN1987A's
equatorial ring has been discovered in the Milky Way: the ring around
the blue supergiant Sher 25 in NGC~3603 (Brandner et al.\ 1997).  More
recently, Smith (2007) has suggested another twin of SN1987A's nebula
around the Galactic LBV candidate HD168625.  Two other notable
examples are the ringed planetary nebulae WeBo1 (Bond et al.\ 2003)
and SuWt2 (Schuster \& West 1976).  Regardless of the nature of their
central stars, these objects are similar in the unusually distinct
ring morphology of their ejecta.  To this class we add two more
examples, both discovered in the Carina Nebula (NGC 3372; see Fig.\
1).

\section{OBSERVATIONS}

\subsection{CTIO Mosaic Images}

The discovery of the two ring nebulae described in this paper was made
serendipitously during a narrowband imaging survey of the Carina
Nebula, although these images are not shown here as they are of lesser
quality than the Magellan images described below.  We surveyed the
entire 2$\fdg$5$\times$4\arcdeg\ area of the giant Carina Nebula on
2003 March 10 using several pointings with the 8192$\times$8192 pixel
imager MOSAIC2 mounted at the prime focus of the CTIO 4m Blanco
telescope.  This camera has a large 35$\farcm$4 field of view, which
allowed us to efficiently survey the whole nebula using the [O~{\sc
iii}] $\lambda$5007, H$\alpha$+[N~{\sc ii}], and [S~{\sc ii}]
$\lambda\lambda$6717,6731 narrowband filters, plus the $i$' broadband
filter.  Further details of the observations and data reduction can be
found in previous papers from the same project (e.g., Smith et al.\
2004, 2005$a$).  Because we were trying to maximize efficient spatial
coverage of the nebula in multiple filters, we did not do a full set
of {\sc mosdither} exposures to correct for the $\sim$3\arcsec\
interchip gaps.  One of the objects of interest here (SBW2) ended up
straddling a gap, motivating the subsequent imaging with the 6.5m
Magellan telescope described below.  Both nebulae were easily
recognized as circumstellar rings in the H$\alpha$+[N~{\sc ii}] filter
images, but no circumstellar emission was seen in images in the other
filters taken with similar 5--10 minute exposure times.

The positions of these two ring nebulae are listed in Table 1, and
their approximate locations are marked on the wide-field image in
Figure 1.  These absolute sky coordinates from the MOSAIC2 images were
determined with reference to USNO catalog stars.

\subsection{Magellan Images}

Following the initial discovery in CTIO images, we obtained narrowband
H$\alpha$ images of the two Carina ring nebulae on 2005 February 21
using the CCD imager MagIC mounted on the 6.5m Clay telescope of the
Magellan Observatory.  MagIC\footnote{see
\tt{http://www.ociw.edu/lco/magellan/instruments/MAGIC/ } } has a
2048$\times$2048 pixel SITe detector, with a spatial pixel scale of
0$\farcs$069 and a 142\arcsec$\times$142\arcsec\ field of view on the
6.5m Clay telescope.  The night was decidedly non-photometric with
patchy and highly-variable cirrus, but the seeing was still quite good
at $\sim$0$\farcs$6.  SBW1 was bright enough to observe easily with a
6.5m telescope looking through clouds, and we took a set of 3
exposures of 120 sec each.  SBW2 is much fainter and while we observed
it the cloud cover became more crippling, but with periodic thin
patches.  Therefore, we took a long series of about 30 consecutive 300
s exposures as the clouds rolled by, and we selected only the
highest-quality images (the ones with the largest number of counts and
best seeing) to shift and add together.  The small 0$\farcs$069 pixels
were rebinned by a factor of 3 to larger pixels of 0$\farcs$207.  The
final H$\alpha$+[N~{\sc ii}] images of the two rings are shown in
Figure 2, while intensity tracings through the middle of each ring are
shown in Figure 3.

\subsection{NTT/EMMI Echelle Spectra}

We used the ESO Multi-Mode Instrument (EMMI) on the NTT to obtain high
resolution spectra of SBW1 and SBW2 on the nights of 2003 March 11 and
12, respectively (one and two nights after their discovery in CTIO
images).  For both targets we used the number 10 echelle grating and
EMMI's red CCD with a 4096$\times$4096 pixel MIT/LL detector, a pixel
scale of 0$\farcs$166$\times$0.058 \AA, and
$R$=$\lambda$/$\Delta\lambda$=28,000 (10 km s$^{-1}$) at wavelengths
near H$\alpha$.  We calibrated the wavelengths using an internal
emission lamp.

For SBW1, we used EMMI in cross-dispersed mode, with a
15\arcsec$\times$1\arcsec\ slit aperture, oriented at P.A.=45\arcdeg\
across the minor axis of the nebula (see Fig.\ 4$a$).  SBW1 is in a
region of bright nebular emission in the western Carina Nebula, so we
also obtained similar exposures of an adjacent position offset by
30\arcsec\ SE to subtract emission from the nebula and sky.  The red
[S~{\sc ii}] $\lambda\lambda$6717,6731 doublet was faintly visible in
the resulting spectra, but the signal to noise in the 30-minute
exposure was too low to provide a meaninful measure of the electron
density.  This is consistent with the fact that we detected no
extended structure in narrowband [S~{\sc ii}] images.  Aside from
H$\alpha$ and [N~{\sc ii}], no other emission lines were detected in
the nebula.  The order including H$\alpha$+[N~{\sc ii}] emission is
shown in Figure 5$a$, while tracings of the spectrum through the NE
and SW portions of the ring are shown in Figure 5$b$, and an extracted
spectrum of the central star is shown in Figure 5$c$.  Detailed views
of the long-slit kinematics for H$\alpha$ and [N~{\sc ii}]
$\lambda$6583 are shown in Figures 6$a$ and $b$, respectively.

For SBW2, which has a larger size on the sky, we used a longer
6\arcmin$\times$1\arcsec\ aperture, with a flat mirror in the beam
instead of a cross-disperser, and we used a narrowband order-sorting
filter to isolate H$\alpha$+[N~{\sc ii}].  This long slit was also
oriented at P.A.=45\arcdeg\ through the minor axis of the ellipse
(Fig.\ 4$b$).  Unfortunately, SBW2 was observed at the end of the
night and had less than half much time as we would have liked, an the
resulting signal to noise of the spectrum was much less than half of
what this object deserved.  The resulting spectrum is noisy and is not
displayed here (dominated by read noise, since it is in a region of
very faint nebular background emission; see Fig.\ 1).  These spectra
are useful, however, because although we detected the
H$\alpha$+[N~{\sc ii}] lines, we detected no Doppler shifts from one
side of the ring to the next.  The limiting resolution of 10 km
s$^{-1}$ for these data then imply a limiting intrinsic radial
expansion speed of $\pm$8 km s$^{-1}$ ($\pm$16 km s$^{-1}$ across the
diameter of the ring) when corrected for the disk inclination angle
(see below).

\subsection{CTIO 4m/RC Spec Spectra}

We used RC Spec on the CTIO 4m to obtain medium-resolution
(R$\approx$6,000) red spectra of SBW1, SBW2, and SuWt2 on 2006 March
15.  RC Spec\footnote{see {\tt
http://www.ctio.edu/spectrographs/4m\_R-C/4m\_R-C.html}.} is the
cassegrain grating spectrograph on the V.M.\ Blanco 4m telescope,
equipped with a Loral 1k$\times$3k CCD detector.  We used grating G380
and filter GG495 to sample wavelengths $\Delta\lambda$=5480--7153 \AA,
with a pixel scale of 0$\farcs$5$\times$0.55 \AA, and an effective
2-pixel spectral resolution of roughly 50 km s$^{-1}$.  The slit width
was set at 0$\farcs$8, matched approximately to the seeing, with the
long slit oriented as in Figure 4 for SBW1 and SBW2, and along the
major axis of the ring at P.A.=--45\arcdeg\ for SuWt2.  We used total
integration times of 9 min for SBW1, and 40 min for SBW2 and SuWt2.
The conditions were photometric, and the data were flux calibrated
using observations of the spectrophotometric standards LTT3218 and
LTT4816.  Wavelength calibration was performed with reference to an
internal HeNeAr lamp.  Emission from the background sky and H~{\sc ii}
region was subtracted by sampling nebular emission along the slit
adjacent to each object.

The extracted flux-calibrated spectra are shown in Figure 7, and line
intensities for several emission lines relative to H$\alpha$=100 are
listed in Table 2.  The relative line intensities in Table 2 have been
dereddened.

A reddening of $E(B-V)$=0.5 for SBW2 is a standard value for
relatively unobscured regions in Carina (e.g., Smith et al.\ 2005b),
while we chose a higher value of $E(B-V)$=1.0 for SBW1, even though it
too appears in Carina, for reasons justified later in \S 3.1.2.  The
value of $E(B-V)$=0.25 for SuWt2 was determined by matching the
continuum shape of the central star in our spectra to an early A-type
photosphere (however, this value is very uncertain due to the small
overall wavelength range of our red spectra).  Uncertainties are a few
percent for the brightest lines, and several to 10\% for the faintest
lines.  For SBW1 and SBW2 we give upper limits to the intensity of
[N~{\sc ii}] $\lambda$5755, since it is important for constraining the
electron temperature and chemical abundances.

On 2006 March 16 we also obtained a blue spectrum of the central star
of SBW1 with RC Spec on the CTIO 4m telescope in order to determine
the spectral type of the central star.  Two 150-second exposures
covering roughly 3000--6000 \AA\ were obtained with the 0$\farcs$67
slit along a position angle of about 10\arcdeg.  The data were reduced
in the standard way, and emission from the sky and background H~{\sc
ii} region was subtracted by fitting background diffuse emission along
the slit.  Figure 8 shows a normalized spectrum extracted from a
$\sim$1$\farcs$5 segment of the slit.

\subsection{New observations of SuWt2 and WeBo1}

Using the MOSAIC2 camera on the CTIO 4m telescope (during the same
survey of star-forming regions when we discovered SBW1 and SBW2 as
noted earlier in \S 2.1), we also obtained an H$\alpha$+[N~{\sc ii}]
image of another ring nebula called SuWt2 on the same night of 2003
March 10.  SuWt2 was originally discovered on red photographic plates
by Schuster \& West (1976), but the new CCD image presented here
(Fig.\ 9) has greater sensitivity and was obtained in better seeing
conditions with a narrow filter.  The new image of SuWt2 will be
discussed in \S 5.

The ring nebula WeBo1 was first reported by Bond et al.\ (2003), who
discussed its morphology in images and the properties of the central
star.  It was discovered independently on 1999 October 6 during our
narrow-band imaging survey of nearby star-forming molecular clouds,
seen on a wide-field narrow-band H$\alpha$\ image obtained with the
1\arcdeg\ field-of-view MOSAIC CCD camera mounted on the Kitt Peak 0.9
meter telescope.  This prompted us to obtain additional imaging and
spectroscopy of WeBo1.

A deep H$\alpha$\ image was obtained at the f/10 Nasmyth focus of the
Astrophysical Research Consortium (ARC) 3.5 meter telescope at the
Apache Point Observatory near Sunspot, New Mexico, on 2001 September
25 using a 2048 $\times$ 2048 pixel CCD (SPICam) with a 40 \AA\
passband H$\alpha$\ filter and a total exposure time of 1160 seconds.
The CCD was binned 2$\times$2 to provide a pixel scale of 0$\farcs$22.
This image is shown in Figure 10$a$.  Low dispersion (R = 500) spectra
were then obtained on 2001 September 26 with the Dual-channel Imaging
Spectrometer (DIS) at the 3.5 meter telescope, using a 0$\farcs$9 slit
width.  The slit was oriented along the major axis of the ring as
indicated in Figure 10$a$, and included the light of the two brightest
stars located in the ring interior.  The resulting spectrum of the
northern portion of the ring is shown in Figure 10$b$.  Additionally,
a high-resolution (R = 20,000) spectrum, shown in Figure 10$c$, was
obtained on 2000 December 18 with the HIRES spectrometer on the Keck I
telescope.  The aperture traced the northern edge of the nebula with
the slit aligned along the major axis of the ring.

Finally, we searched for J = 1--0 CO emission in a 10\arcmin\ diameter
box using the 16 beam SEQUOIA array at the 14 m diameter radio
telescope at the Five College Radio Astronomy Observatory (FCRAO) near
Amherst Massachusetts on 2001 January 6.  No emission was found at any
velocity above an rms brightness temperature level of 0.3 K in 92 kHz
wide channels.

\section{RESULTS}
\subsection{Imaging and Spectroscopy of SBW1}

\subsubsection{Basic Morphology and Kinematics}

SBW1 appears as a distinct, almost perfectly smooth elliptical nebula
with a well-defined outer edge in Figure 2$a$. With its bright central
star it resembles a wide-open staring eye, lidless and wreathed in
flame, probably deserving a nickname like the ``Eye of Sauron''
nebula; later in this paper we will discuss its relation to the other
rings.  This ring is brightest at its southeast and northwest apexes,
due to limb-brightening, and the interior of the ring is partially
filled-in with diffuse emission. Figure 3$a$ shows an intensity
tracing through the minor axis of the nebula, along with a model of
the expected emission for a slice through a limb-brightened shell
where the inner and outer radii listed in the figure give the best fit
to the intensity of the edges of the nebula.  Clearly it is plausible
that additional emission fills its center --- however, the kinematics
of this nebula are unusual, as discussed below.

The outer limb-brightened edge of the equatorial ring allows us to
measure its geometric parameters accurately.  The major axis of the
ring is 11$\farcs$1$\pm$0$\farcs$1, while the minor axis is
7$\farcs$1$\pm$0$\farcs$1.  This yields an inclination of
$i$=50$\fdg$2$\pm$1\arcdeg\ if the ring is circular.  The polar axis
(corresponding to the direction of the minor axis on the sky) lies at
P.A.=32$\fdg$4$\pm$1\arcdeg.  The intrinsic physical dimensions of the
ring depend on the distance, which turns out to be somewhat surprising
as discussed in \S 3.1.2.

Faint but visible in Figure 2$a$ are a set of outer rings, another few
arcseconds to the northeast and southwest.  These remind one of the
fainter outer rings around SN1987A, although they do not have the same
special magnificence, appearing very thin, like butter scraped over
too much bread.  They give the impression of limb-brightened edges of
faint bipolar lobes above and below the equatorial plane defined by
the brighter inclined ring.  In fact, the image of SBW1 in Figure 2$a$
resembles the SN1987A model image of Martin \& Arnett (1995) more
closely than SN1987A itself does.  This might lead one to suspect that
the true geometry of SBW1 is akin to that of the bipolar model of
Martin \& Arnett.  An even more extended, and fainter third set of
rings or polar lobes may be present as well --- these are hard to see
on the printed page because of the grayscale levels in Figure 2$a$,
but their emission can be seen at $\pm$9\arcsec\ from the star in the
intensity tracing along the polar axis of the nebula in Figure 3$a$.
Deeper images may help reveal these putative outer rings, although
they are nearly as faint as the fluctuations in the background nebula.

High-resolution spectra of SBW1 in Figures 5 and 6 clearly show
extended nebular H$\alpha$ and [N~{\sc ii}] emission lines.  The
detailed kinematic structure of the emission along the slit shown in
Figure 6 is rather bizarre.  The strongest emission at the position of
the main ring is redshifted to the northeast, and blueshifted to the
southwest.  The faint emission that fills the ring's interior in
images follows the same general trend, but the detailed kinematic
structure shows a triangular shape above and below the star in Figure
6.  The emission does not converge on the star at the central
position, as would be expected for a Hubble flow (homologous
expansion), but rather, meets the position of the star at $\pm$12 km
s$^{-1}$.  We offer no obvious explanation for this strange kinematic
structure, except that it is marginally consistent with the interior
of the ring being filled with a steadily outflowing flared disk.  The
cause of the triangular kinematic shape may be geometric, such that
the motion of material was initially ejected or subsequently diverted
out of the equatorial plane to form a flaring ring.  In that case, the
centroid of this emission at $\pm$12 km s$^{-1}$ implies a radial
expansion speed for the ring and disk of roughly 19$\pm$2 km s$^{-1}$
when it is corrected for the inclination angle of 50$\fdg$2.  This
expansion speed is identical to the radial expansion speeds of the
rings around the blue supergiants Sher 25 and HD168625.

\subsubsection{The central star and its distance}

Figure 8 shows the spectrum of the central star of SBW1, which
exhibits a spectral type of B1.5 Iab.  The luminosity class is not
entirely clear, because higher dispersion (Fig.\ 5) shows that the
Balmer absorption lines are partially filled-in with nebular emission.
The extracted spectrum of the bright central star in Figure 5$c$ shows
some nebular emission rising above the continuum level in H$\alpha$
and double-peaked in the [N~{\sc ii}] $\lambda$6583 line, superposed
on a smooth continuum spectrum with prominent broad Balmer absorption.
Despite the nebular contamination, a spectral type of B1.5 Iab seems
likely.  The luminosity class has two important ramifications
concerning implications for stellar evolution, as well as the distance
to SBW1.

First, it means that the central star is a massive star with an
initial mass of roughly 18--25 M$_{\odot}$, and the ring is therefore
an example of a pre-supernova environment much like that around
SN1987A.  The spectral type of B1.5 Iab justifies a close comparison
to both Sher 25 (B1.5 Ia; Moffat 1983) and the progenitor of SN1987A
(B3 I; Walborn et al.\ 1989), as well as permitting a link to the B[e]
supergiants because of the emission line spectrum (see \S 4.2).

Second, the high luminosity suggests a distance much larger than 2.3
kpc, meaning that it is seen through the Carina Nebula and is
projected within its boundaries only by chance, as if it
wants to be found there.  The absolute visual magnitude of a B1.5 Iab
supergiant should be about $-$7.0 (Crowther et al.\ 2006), and the
observed visual magnitude is roughly m$_V$=12.7 (Table 1).  If SBW1
were located at 2.3 kpc in the Carina Nebula, the required total
line-of-sight extinction would be very large at A$_V\simeq$7.9
magnitudes.  Such high extinction is plausible, at least in principle,
since SBW1 is seen along the same sight-line as dense molecular gas
and dust associated with the Carina Nebula (Zhang et al.\ 2001;
Grabelsky et al.\ 1988; Smith et al.\ 2000).  However, the 2MASS
colors ($J$=10.55, $H$=10.18, and $K$=9.95) suggest that the visual
extinction is more like 3-5 magnitudes, depending on the extinction
law.  This implies that the star appears faint for its luminosity
class because it is actually at a much farther distance.

In fact, the observed radial velocity of SBW1 also implies a much
larger kinematic distance that places it in the far Carina arm.  The
systemic velocity in Figure 6 is redshifted from the centroid of
nebular emission in Carina by roughly +30 km s$^{-1}$.  The
heliocentric radial velocity of $\eta$ Carinae and the Carina Nebula
is --8.1 km s$^{-1}$ (Smith 2004), but that of SBW1 is +22$\pm$4 km
s$^{-1}$, or an LSR velocity of roughly 10$\pm$4 km s$^{-1}$.  The
Carina Nebula is seen near the tangent of the Sagittarius-Carina
spiral arm (e.g., Bok 1959), and a positive radial velocity at its
position places SBW1 outside the solar circle on the far side of the
Milky Way.  The corresponding distance at a longitude of
$l$=287\arcdeg\ for a flat rotation curve is $\sim$7 kpc.\footnote{But
we may yet be deceived.  There is some hint in our [O~{\sc iii}]
images and long slit spectra that the ring may be seen in silhouette
against the background nebular emission.  If true, this would mean
that SBW1 must be located within the Carina Nebula after all and that
its radial velocity is anomalous, unless there is another H~{\sc ii}
region projected behind Carina.  Higher signal to noise data of the
same type or possibly UV spectroscopy of absorption line profiles to
the star may resolve this ambiguity.  If SBW1 is far beyond Carina,
the far (receding) side of the H~{\sc ii} region should be seen in
absorption.  Thus, there are still questions that need answering.}

This larger distance joins SBW1 to a string of other famous hot
supergiants with circumstellar nebulae in the far Carina arm, seen on
either side of the Carina Nebula, such as Sher~25 in NGC3603,
AG~Carinae, Hen~3-519, and HR~Carinae.  The massive cluster Westerlund
1 is also thought to be at a similar distance in this part of the
Galactic plane.  SBW1 is less conspicuous than these stars and its
ring has not been discovered until now because of its much higher
extinction column along its line of sight through dust and gas in the
Carina Nebula, or perhaps, simply because it is overshadowed by other
spectacular objects in the brighter inner parts of the nebula.  At 7
kpc, the required extinction for its apparent magnitude is roughly 5
magnitudes.  This is in much better agreement with the extinction
implied by near-IR photometry as noted above.  Also, the larger
distance of 7 kpc means that the ring's radius is about 0.19 pc ---
almost identical to the rings around SN1987A and Sher~25.  With a
radial expansion speed of 19 km s$^{-1}$, its dynamical age is
$\sim$10$^4$ yr.  This age, too, is comparable to the rings around
SN1987A and Sher~25.

Its high luminosity and circumstellar ring make SBW1 an object of
considerable interest, and its similarity to the equatorial ring
around SN1987A hints at some things which have not yet come to pass.
The central star deserves further observations and analysis, such as
monitoring for photometric and spectroscopic variability (i.e. is it
an eclipsing or spectroscopic binary?), as well as detailed
atmospheric spectral analysis to accurately determine its physical
parameters.  This is important if SBW1 is a potential supernova
progenitor.

\subsubsection{Size, mass, and chemical abundances}

Table 2 lists line intensities for SBW1, corrected for a reddening
value of $E(B-V)$=1 that would correspond to our estimate of
$A_V\simeq$5, adopting the value of $R$=4.8 that is appropriate for
extinction in the direction of Carina (e.g., Smith 2002).  From the
red [S~{\sc ii}] $\lambda\lambda$6717,6731 doublet, we measure an
electron density of n$_e$=513 cm$^{-3}$, and from the ratio [N~{\sc
ii}] $\lambda$6548+$\lambda$6583/$\lambda$5755 we find an upper limit
to the electron temperature of T$_e <$12,350 K.  In the extracted EMMI
spectrum of the nebula (Fig.\ 5$b$) we measure a [N~{\sc ii}]
$\lambda$6583/H$\alpha$ intensity ratio of 0.483, which is within a
few percent of the value we measure from RC Spec data.  This would
yield a N$^+$/H$^+$ abundance of $>$1.6$\times$10$^{-5}$ for an upper
limit of T$_e<$12,350 K, or $N^+/H^+\simeq$2.7$\times$10$^{-5}$ for a
more likely temperature of T$_e$=10,000 K.  This is still {\it lower}
than the solar abundance level; the nitrogen abundance would be
roughly solar if the electron temperature were as low as 7,000 K
(note, however, that this is only the abundance of N$^+$).  The
electron temperature would need to be roughly 5000 K or less, or there
would need to be a large neutral N fraction compared to hydrogen in
order for this ratio to suggest a considerable N overabundance in the
ring.  Therefore, our data are consistent with no significant N
enrichment in the ring of SBW1.  This lack of distinct nitrogen
enrichment implies that SBW1 {\it has not yet passed through a red
supergiant phase}, much like its twin Sher 25 (Smartt et al.\ 2002).
Thus, in both these cases the equatorial rings must have been ejected
when the stars were blue supergiants.  This has important implications
for the rings around SN1987A (Smith 2007; Smith \& Townsend 2007).

If the ring is fully ionized, we can estimate its mass from the
electron density and its apparent geometry in images.  If we
approximate the geometry of the ring as a torus, then at a distance of
7 kpc, the ring's radius is $R\simeq$0.19 pc or 5.9$\times$10$^{17}$
cm, and the radius of its cross-sectional area is
$r\simeq$5.3$\times$10$^{16}$ cm ($\sim$0$\farcs$5). Then assuming
pure H gas, the ring's mass would be

\begin{equation}
M \simeq 2 \pi^2 r^2 R \, m_H \, n_e \, . 
\end{equation}

\noindent With n$_e$=513 cm$^{-3}$ (Table 2), we find M$\approx$0.014
M$_{\odot}$.  This is probably a lower limit to the total mass,
however, because we have neglected any neutral fraction or dense
clumps, and we have ignored the mass of the disk interior to the ring,
which may be comparable to the ring itself.  This mass is comparable
to the range of masses inferred for the few hundred AU disks around
B[e] supergiants (Zickgraf et al.\ 1986).  The largest mass known in a
B[e] star disk is about 0.3 M$_{\odot}$ around R126 (Kastner et al.\
2006).  It is also comparable to the mass for the rings around RY
Scuti (Smith et al.\ 2002b).

It would be interesting to measure the thermal-infrared flux from
SBW1, in order to measure its dust mass.  Our long-slit spectra
obtained with RC Spec hint that significant dust may be present in the
nebula, since it is apparently seen in silhouette, blocking light from
the background nebula in some emission lines (the low signal to noise
and foreground emission preclude an accurate estimate of the dust
mass, however).

\subsection{Imaging and Spectroscopy of SBW2}

The structure of SBW2 is far less orderly than SBW1.  It is much
fainter, more diffuse and ratty-looking, less symmetric, and it lacks
a clear outer edge.  The most unnatural and curious property, however,
is that there is no obvious bright star near its center.  There are
three faint stars inside the ring (with $R\simeq$18.5, 20.5, and 20.7
mag), as well as a few extremely faint stars that are scarcely
detected in our Magellan image with $R$ of about 26 to 27.  However,
none of these stars is at the ring's center, and none stands out as an
obvious candidate for a central star that may have ejected it.  The
sound crossing time of the ring is only a few tens of thousands of
years.  Any white dwarf at that young age should be several magnitudes
brighter at visual wavelengths than these stars (e.g., Bergeron et
al.\ 1995) if SBW2 is within the Carina Nebula at 2.3 kpc.  Thus, the
central star in SBW2 might avoid being seen if it were older, but to
disappear entirely, that is a rare gift.  Is the central object
completely obscured by its own dust?  Probably not, as we should see
the outside of such an envelope ionized by whatever ionizes the rest
of the ring.  Perhaps the most likely -- if unsatisfying --
explanation is that a tight binary system was disrupted when one of
the stars shed a large fraction of the system's mass.  The origin of
the ring and its invisible star system has since passed out of all
knowledge, but proper motions of stars in the surrounding field may be
quite illuminating.

The purpose of the intensity tracing in Figure 3$b$ is to illustrate
that SBW2 is not a limb-brightened spheroidal shell, but an actual
flattened ring with a cleared interior.  The intensity tracing through
the minor axis of the nebula shows that the brightness in the middle
of the ring drops to roughly zero.  On the other hand, the dotted line
in Figure 3$b$ shows a model of the brightness distribution we would
expect for a limb-brightened shell with a thickness corresponding to
the apparent ring thickness in images.  It is clear that the central
region of the nebula would be more filled-in if it were a
limb-brightened shell, unless the ring were much thinner than
observed.  Thus, the morphology of SBW2 is that of a flattened
equatorial ring with a hollow center, titled at an inclination angle
of about 52\arcdeg.  The average radius of the ring is 19$\farcs$8, or
0.22 pc if it is in the Carina Nebula at 2.3 kpc.  This makes its
average radius only 10\% larger than the equatorial ring around
SN1987A (the inner and outer radii of SBW2 are roughly 0.16 and 0.25
pc, respectively).

Another unusual or unexpected property of SBW2 is that the brightest
parts of the ring are not at the southeast and northwest tangent
points along the major axis, as would be expected from
limb-brightening in an optically-thin ring like SBW1.  In fact, the
brightness distribution is the opposite --- the brightest parts of
SBW2 are the near and far sides along the {\it minor} axis.  This
points to a high degree of azimuthal asymmetry in the system, perhaps
implying that an eccentric binary system played a role in shaping the
ejecta.

Although SBW2 is projected within the boundaries of the star-forming
Carina Nebula, it is unlikely that this is a normal pre-main-sequence
circumstellar envelope.  In our experience, no pre-MS stars or
protostars are surrounded by circumstellar disks that look anything
like SBW2. They either have a bright central star with a bipolar
reflection nebula or jets, or if the central star is obscured, it is
because it is seen behind a dark, edge-on obscuring lane like in HH30
(Burrows et al.\ 1996; Stapelfeldt et al.\ 1999), HK~Tau (Stapelfeldt
et al.\ 1998), HV~Tau C (Stapelfeldt et al.\ 2003), and several other
reflection nebulae in Taurus and Orion (Padgett et al.\ 1999).  This
is clearly not the case here.  In some situations like debris disks
around Vega-like stars, images of the circumstellar disks may resemble
SBW2, but they all have bright central stars that overwhelm the nebula
in optical images (e.g., Schneider et al.\ 1999; Fukagawa et al.\
2004; Greaves et al.\ 1998).

This speculation is borne-out when we consider the spectrum of SBW2,
and the chemical abundances it implies.  The spectrum is shown in
Figure 7$b$, with line intensities listed in Table 2. The strong
[N~{\sc ii}] lines relative to H$\alpha$ immediately raise suspicion
that the gas is nitrogen rich.  (It also means that our H$\alpha$
image in Figure 2$b$ is actually a [N~{\sc ii}] image.)  With n$_e$ of
$\sim$280 cm$^{-3}$ and an upper limit of T$_e <$14,900 K (Table 2),
we find a lower limit of n(N$^+$)/n(H$^+$)$>$1.35$\times$10$^{-4}$, or
about 1.5 times the solar abundance of N.  For a more likely
temperature of T$_e \la$10,000 K, the nitrogen abundance would be
$\ga$4 times solar, and this is only for N$^+$.  Thus, the gas in SBW2
is clearly nitrogen rich, and it is therefore the result of
post--main-sequence mass loss, probably from an intermediate-mass
progenitor star.  Since the Carina Nebula H~{\sc ii} region is only
about 3 Myr old, this object must be an interloper, passing through
the nebula instead of having been born there.  The lack of any
ionizing source within the ring argues strongly that it is within the
Carina Nebula, and not just projected along the same sightline, like
SBW1.

The emitting layer thickness of the ring seems to be roughly
$l\simeq$2$\times$10$^{17}$ cm (corresponding to a projected width of
$\sim$5\arcsec; see Fig.\ 2$b$).  With a density of n$_e$=280
cm$^{-3}$ indicated by the [S~{\sc ii}] lines (Table 2), it requires
an incident flux of ionizing photons of $l \alpha_B n_e^2 > 10^{9.6}$
s$^{-1}$ cm$^{-2}$ in order to be fully ionized
($\alpha_B$=2.6$\times$10$^{-13}$ cm$^3$ s$^{-1}$ is the case B
recombination coefficient).  At a projected radius of 70\arcmin\ or
$\sim$45 pc from the central ionizing clusters of the Carina Nebula,
the ambient Lyman continuum flux near SBW2 would be roughly $Q_H / (4
\pi d^2)\simeq$10$^{9.6}$ s$^{-1}$ cm$^{-2}$ as well (Smith 2006).
Thus, the star clusters of the Carina Nebula supply sufficient
radiation to keep SBW2 fully ionized.  This argues that SBW2 is, in
fact, located within the Carina Nebula itself and not just projected
along the same sightline by chance.  If our assumption that SBW2 is
fully ionized is incorrect, and that only its outer surface is
ionized, then the required ionizing flux goes down.  While the nebula
is ionized, it lacks detectable [O~{\sc iii}] emission in our images
(not shown).

Assuming that SBW2 is fully ionized, we can provide a lower limit to
its mass using the same technique as we did for SBW1.  Using equation
(1), but with $r$=1.7$\times$10$^{17}$ cm (4$\farcs$9) and
$R$=6.6$\times$10$^{17}$ cm (0.22 pc) from Figure 2$b$, and n$_e$=280
cm$^{-3}$ from Table 2, we find a total mass of $\sim$0.1~M$_{\odot}$
of ionized gas.

\subsection{Imaging and Spectroscopy of SuWt2}

This exotic ring nebula is shown in Figure 9$a$, which is an
H$\alpha$+[N~{\sc ii}] image obtained with the MOSAIC2 camera on the
CTIO 4m telescope.  This image shows considerably more detail than the
discovery plates (Schuster \& West 1976).  From the image in Figure
9$a$ we measure an inclination angle of 64\arcdeg($\pm$2\arcdeg), and
from Figure 9$b$ we find an average radius of 40$\farcs$5 or about 0.2
pc at a distance of 1 kpc, making it yet another ring with roughly the
same radius as that of SN1987A's ring.  As was the case for SBW2, an
intensity tracing across SuWt2 (Figure 9$b$) shows that it is not a
limb-brightened ellipsoidal shell, but instead is an actual flattended
equatorial ring.  A shell model would require substantially more flux
in the center of the nebula than is seen in this image.  Rather than
being partially filled in, SuWt2 has faint excess flux {\it outside}
the ring, suggesting that the ring may be the inner edge of a swept-up
disk.  The central star of SuWt2 has been found to be a 4.91-day
eclipsing binary composed of 2 A-type stars of roughly 2.5 M$_{\odot}$
each (Bond 2000; Bond et al.\ 2002), where the inclination is in rough
agreement with our estimate of the inclination of the ring nebula.

Figure 7c shows the red spectrum of the ring nebula, and Table 2 lists
measured line intensities corrected for $E(B-V)$=0.25.  We do not have
a reliable estimate of $E(B-V)$ for SuWt2, but this reddening value is
roughly what is needed to make the observed stellar colors consistent
with an A-type photosphere.  The strong nitrogen lines indicate that
the H$\alpha$+[N~{\sc ii}] image is dominated by [N~{\sc ii}]
$\lambda$6583 emission, and hint at an enhanced N abundance.  With
n$_e\simeq$100 cm$^{-3}$ (the observed [S~{\sc ii}] lines are close to
the low density limit) and T$_e$=11,400 K from the [N~{\sc ii}] line
ratio, the observed [N~{\sc ii}] $\lambda$6583/H$\alpha$ ratio
indicates n(N$^+$)/n(H$^+$)$\simeq$2.5$\times$10$^{-4}$, or almost 3
times the solar abundance of N. Again, this is only for N$^+$, so the
true N abundance could be higher.  The presence of nitrogen-rich
ejecta confirms that SuWt2 is a post--main-sequence object, and
suggests that it had an intermediate-mass progenitor star.  This is
curious, given that the central star system of SuWt2 is observed to be
an eclipsing binary of two A-type main-sequence stars.

Except for their central stars, SuWt2 and SBW2 are near twins.  Again,
we find that the ionization of this ring is rather suspicious.
Ionized gas in the ring nebula requires a stronger flux of ionizing
photons than can be supplied by two A-type stars.  Bond et al.\ (2002)
hypothesized that there may be a third star in the system --- probably
an unseen white dwarf --- to account for the necessary UV flux.
However, this white dwarf would need to be in a wide orbit, detached
from the A+A eclipsing binary, and would therefore not be able to
strongly affect the shape of the ring.

However, here we note a speculative alternative.  Figure 9 also shows
a bright star about 1\arcmin\ east/northeast of SuWt2.  This is the B2
star SAO 241302.  It could supply the required ionizing photons if the
two are at the same distance, following the same reasoning as above
for SBW2 in Carina.  This may seem at first to be rather fortuitous.
However, we note that the H$\alpha$ image of SuWt2's ring in Figure 9
shows an asymmetric brightness distribution.  In particular, the ring
is brighter on the east/northeast side --- exactly on the side facing
the bright B star.  Thus, much like SBW2 in Carina and possibly also
Sher 25, the nebula may be {\it externally illuminated} by a nearby
hotter star.  This possibility requires further investigation.

\subsection{Imaging and Spectroscopy of WeBo1}

This ring-shaped planetary nebula is located near the star forming
region W4 in the Perseus arm of our Galaxy, about 5\arcmin\ southwest
of the X-ray binary LS I +61\arcdeg303 (WeBo1 was discovered
serendipitously while studying this source; Bond et al.\ 2003).  It
surrounds a Barium star and suspected white dwarf that may be the
source of ionizing photons for the nebula.  Bond et al.\ (2003)
proposed that the equatorial ring was ejected during the
tidally-locked AGB phase of the more massive star in the binary before
it became a white dwarf, and that some of the AGB wind was accreted
onto the companion star that is seen today as a rapidly-rotating
Barium star.  Thus, while the star shows no eclipses or orbital reflex
motion that have been detected yet, the atmospheric chemical
abundances give a strong indication that close binary evolution played
a role in the axisymmetric mass loss.

\subsubsection{New Results}

Figure 10$a$ shows an H$\alpha$\ image of the ring obtained with the
ARC 3.5 meter reflector.  The ring has a sharp inner edge with a
radius of about 25\arcsec.  As noted already by Bond et al.\ (2003),
the lack of emission in the ring's interior means that it is a toroid
and not a limb-brightened ellipsoidal shell.  The major to minor axis
ratio of 3.85 implies that the ring is close to edge-on with an axis
of symmetry inclined by 75\arcdeg$\pm$3\arcdeg\ with respect to the
line of sight.  In contrast to the sharp inner boundary, the outer
edge of the ring is diffuse, much like SBW2 and SuWt2.

The line intensities in Table 2 are measured from the spectra in
Figures 10$b$ and $c$.  The higher resolution Keck spectrum was used
for the relative intensities of [N~{\sc ii}] $\lambda$6548 and
$\lambda$6583 because they were better separated from H$\alpha$ than
in the lower resolution DIS spectrum.  The observed line intensities
have been corrected for $E(B-V)$=0.57 (Bond et al.\ 2003).  Table 2
also lists the electron density, electron temperature, and N$^+$/H
abundance calculated from these ratios.  The red [S~{\sc ii}] lines
are unusually weak in the spectrum, and unfortunately, the
corresponding larger uncertainty in their intensities introduces a
factor of $\sim$2 uncertainty in the electron density.  The critical
[N~{\sc ii}] $\lambda$5755 line is also very weak, with a marginal
2$\sigma$ detection that we regard as an upper limit.  This indicates
an upper limit to the electron temperature of $\la$10$^4$~K, which
seems reasonable enough.  From these values and the strength of
[N~{\sc ii}] $\lambda$6583, we find a N$^+$/H$^+$$\ga$10$^{-4}$,
indicating that the nebular gas is strongly enhanced in nitrogen, well
above the solar value.  This points toward an intermediate mass AGB
progenitor star, just as in the case of SBW2 and SuWt2.

Interestingly, we see weak emission from He~{\sc i} $\lambda$5876 and
[Ar~{\sc iii}] $\lambda$7136 in the spectrum.  These high ionization
lines cannot be caused by radiation from the cool central star that
dominates the visual spectrum, requiring either a hot white dwarf or
an external source of ionization.  We note that WeBo1 is found in the
outskirts of the giant H~{\sc ii} region W4, and is near the famous
microquasar and radio emitting X-ray binary LS I +61\arcdeg 303.  This
proximity is intriguing, although the morphology in the [O~{\sc iii}]
image presented by Bond et al.\ (2003) seems to suggest a
centrally-located ionizing source.

\section{DISCUSSION}

\subsection{Comparison with Other Rings}

{\bf SN1987A:} Probably the most famous equatorial ring associated
with post--main-sequence evolution is the remarkable equatorial ring
around SN1987A.  It was discovered shortly after being flash-ionized
by the supernova, and eventually became the subject of intense study
with {\it HST} (Crotts, Kunkel, \& McCarthy 1989; Lundqvist \&
Fransson 1991; Jakobsen et al.\ 1991; Plait et al.\ 1995; Burrows et
al.\ 1995; etc.).  Now is a particularly interesting time for study of
SN1987A, as the supernova blast wave is overtaking the circumstellar
ring and causing a host of phenomena that are of great interest to
shock evolution (Luo \& McCray 1991; Chevalier \& Dwarkadas 1995;
Borkowski et al.\ 1997; Michael et al.\ 2000; Pun et al.\ 2002; Smith
et al.\ 2005$b$).  The appearance of a number of hotspots around the
ring (Sonneborn et al.\ 1998; Michael et al.\ 2000; Pun et al.\ 2002;
Sugerman et al.\ 2002) make it look like a pearl necklace.  These
spots show that while the ring appeared fairly smooth in images before
it was struck by the shock, it had actually been shaped by
Rayleigh-Taylor instabilities (Michael et al.\ 2000).  The nebula
around SN1987A is usually assumed to have been ejected during a
previous red-supergiant phase and then swept-up by a blue supergiant
wind. The ring itself is thought to be a swept-up thin disk that
formed in the RSG wind as a consequence of close binary activity --
possibly even a merger (Blondin \& Lunqvist 1993; Martin \& Arnett
1995; Morris \& Podsiadlowski 2006; Collins et al.\ 1999).  However,
the larger outer rings that compose the triple-ring system around
SN1987A still seem to defy explanation.  Soker (2002) hypothesizes
that these rings formed from precessing jets, while Morris \&
Podsiadlowski (2006) suggest that they are a natural consequence of a
binary merger event.  In any case, they are bona fide rings and not
just limb-brightened edges of an hourglass (Burrows et al.\ 1995).
Interestingly, our image of SBW1 in Carina looks more like one of
these hourglass nebula simulations intended for SN1987A (see for
example Fig.\ 5 in Martin \& Arnett 1995).

{\bf Sher 25:} The nebula around this B1.5 supergiant in NGC3603 (Sher
1965; Moffat 1983) is often held up as the Milky Way's ``twin'' of
SN1987A's progenitor.  Sher~25 has an equatorial ring that is
reminiscent of the one around SN1987A (Brandner et al.\ 1997).
However, Sher 25 is more luminous than Sk --69\arcdeg202. With
log(L/L$_{\odot}$)=5.9 (Smartt et al.\ 2002), Sher 25 is probably too
luminous to have gone through a red supergiant phase.  In fact, Smartt
et al.\ find that the observed N/O abundance is incompatible with
Sher~25 having been a RSG, and that the nebula was probably ejected as
a blue supergiant.  Furthermore, the spectrum of Sher~25 does not show
the broad lines we would expect from an extremely rapid rotator that
should be the product of a merger event (e.g., Morris \& Podsiadlowski
2006).  Thus, the similarity of its nebula to that around SN1987A may
pose some difficulties for the RSG/merger scenario for the shaping of
SN1987A's nebula.  Our new discovery of SBW1 provides yet another twin
of Sher~25 and SN1987A.  SBW1 has the same radius as the rings around
Sher~25 and SN1987A, and the central star is also a B1.5 supergiant
(although it is less luminous than Sher 25, with a luminosity class of
Iab instead of Ia).  It also lacks the strong N-enrichment required
for post-RSG evolution.

{\bf HD168625:} The equatorial ring around this LBV candidate was
discovered by Hutsem\'{e}kers et al.\ (1994), and has been studied by
several authors since then.  In a recent study (Smith 2007) we
discovered rings in a larger bipolar nebula seen in {\it Spitzer}
images, making the nebula the nearest Galactic analog of the
triple-ring system around SN1987A.  The detection of an analog to
SN1987A's nebula around an LBV is significant, because it provides a
precedent that massive stars can eject rings+bipolar nebulae in LBV
outbursts as blue supergiants rather than as RSGs.  Elsewhere (Smith
2007) we have discussed this object in more detail, along with strong
implications for the progenitor of SN1987A.

{\bf RY Sct:} This is a massive 11-day eclipsing binary system in a
state of overcontact (Milano et al.\ 1981), probably a WR+OB
progenitor.  Its unusual circumstellar nebula was first spatially
resolved as a limb-brightened dust torus in the thermal-IR, where the
compact torus had an apparent radius of only 1\arcsec\ (Gehrz et al.\
1995; 2001).  The inner wall of this dust torus is ionized by the hot
central stars, giving rise to strong radio continuum emission
(Hjellming et al.\ 1973) and a peculiar, high-excitation emission-line
spectrum (Merrill 1928; Smith et al.\ 2002b).  The morphology of this
ionized component of the nebula is remarkable --- in high-resolution
{\it HST} images it appears to be a pair of plane-parallel thin rings
on either side of the equatorial plane (Smith et al.\ 1999).  Proper
motions indicate that the rings were ejected recently during an event
sometime in the 19th century (Smith et al.\ 2001).  The rings have
radial expansion speeds of a little more than 40 km s$^{-1}$ (Smith et
al.\ 2002b).  These ionized double-rings have half the radius of the
larger dust torus, only 0$\farcs$5 or 900 AU at that distance.  The
inclination angle of the rings agrees to within a few degrees with the
inclination angle of the orbit derived from the eclipsing light curve
(Smith et al.\ 2002b; Milano et al.\ 1981), providing strong empirical
evidence that the nebula and binary system share the same equatorial
plane, and that close binary evolution plays an important role in
shaping the nebula.  RY Scuti is one of our best cases of a massive
star caught in the act of shedding mass and angular momentum through
its outer Lagrange point, probably an important step on its way toward
becoming a colliding-wind WR+OB binary system.  It argues strongly
that equatorial rings can be the consequence of close binary
evolution.

{\bf He 2-147:} This equatorial ring nebula surrounds a binary system
that is a symbiotic Mira.  It was probably ejected in a symbiotic nova
event about 300 yr ago, and currently shows a planetary nebula-like
spectrum in the blue, and a cool supergiant spectrum in the red
(Corradi et al.\ 1999, 2000).  It is a close relative of bipolar
symbiotic planetary nebulae like He 2-104.

Other clear examples of equatorial rings that are not as well studied
as these are SuWt3, Abel 14, M1-41, Abel 47, and Lo 18 (see Schwarz et
al.\ 1992).  SuWt3 (Schuster \& West 1976) is a remarkably clean, thin
ring nebula that looks like a perfect ellipse, with $i \simeq$
58\arcdeg\ and $R$=28\arcsec.  The distance to SuWt3 is not known, but
$R$=0.14 pc $\times$ D$_{\rm kpc}$ makes it comparable to the other
rings if the distance is around 1 kpc.  Another intriguing object that
deserves further study is the planetary nebula Abel 14 (Schwarz et
al.\ 1992).  It shows what seems to be a pair of plane-parallel rings
above and below the equatorial plane, plus an outer torus component to
the nebula (see Manchado et al.\ 1996).  The appearance of the nebula
is very similar to the unusual double-ring nebula around RY Scuti (see
above).  Finally, there are still many more bipolar nebulae with less
well-defined equatorial tori or rings at their waists, like $\eta$
Car, He2-104, NGC 2346, NGC7027, MyCn18, and Abel 55.

\subsection{Rings around B supergiants and a link to the B[e] supergiants}

B[e] supergiants are luminous post-main-sequence stars that are
characterized by IR excess emission from dust and prominent
emission-line spectra, including the Balmer emission lines plus
numerous lines of permitted and forbidden Fe$^+$ and other species
(Zickgraf et al.\ 2003; Miroshnichenko et al.\ 2002).  This emission is
thought to arise in an outflowing circumstellar disk within a few
hundred AU of the star (e.g., Zickgraf et al.\ 1986).  Like the
lower-luminosity Be stars, B[e] supergiants are thought to be rapid
rotators.

B[e] supergiants have high luminosities, and occupy a similar area in
the HR diagram as some of the less massive LBVs.  However, unlike the
LBVs, B[e] supergiants typically do not exhibit marked photometric or
spectroscopic variability.  Perhaps they are an earlier phase
immediately before the LBVs, where the star still has sufficient mass
to insulate it from the near-Eddington LBV instability.  There are
some deviations from this trend, however, hinting that B[e]
supergiants and LBVs may be related after all: R4 in the SMC is a B[e]
supergiant that exhibits variability characteristic of the LBVs
(Zickgraf et al.\ 1996).  Interestingly, R4 is very close to HD168625,
SBW1, and the progenitor of SN1987A on the HR diagram, with a presumed
initial mass of $\sim$20 M$_{\odot}$ (Zickgraf et al.\ 1996).  These
similarities taken together with the equatorial ring geometry suggest
a plausible connection between B[e] stars and more extended rings.
Also, R4 demonstrates that even relatively low-luminosity blue
supergiants (low luminosity, at least, compared to extreme LBVs like
$\eta$ Car) can be susceptible to instabilities that cause sporadic
outbursts.  If a significant IR excess from dust were detected in
SBW1, it would support this connection because the B[e] star disks
also contain dust.

We therefore propose a direct link between B[e] supergiants and the
class of massive, early B supergiants surrounded by spatially-resolved
ring nebulae (including SN1987A, SBW1, Sher 25, and HD168625).
Specifically, we suspect that the outflowing disks in B[e] stars that
emit their IR excess and bright-line spectra will continue to expand,
and after a few thousand years will form spatially resolvable
equatorial rings at $\sim$10$^4$ AU from the stars.  This would be
analogous to the disk clearing that is thought to occur in lower
luminosity Be stars (e.g., Meilland et al.\ 2006).  This link, in
turn, would suggest that in the massive ringed stars in our sample,
rapid rotation rather than close binary evolution may shape their
ejecta.  Both Sher 25 and HD168625 have been studied, and no evidence
for binarity has yet been reported, although renewed efforts would be
of great interest.  SBW1 should also be monitored to search for
eclipses or radial velocity variations.  Both Sher 25 and SBW1 are
luminous hot supergiants that probably have {\it not} gone through a
RSG phase, as noted earlier, while HD168625 is a luminous blue
variable (LBV) candidate.  Thus, the most plausible scenario for the
formation of their rings is ejection by a rapidly-rotating star in an
LBV eruption or some related episodic mass loss as blue supergiants
(see Smith 2007; Smith \& Townsend 2007).  In this context, the fact
that their nebulae are near twins of SN1987A is quite suggestive.

It is worth pointing out that the presence of descrete rings -- as
opposed to smooth thin disks -- is evidence for recent temporal
variability of the central star in each object, regardless of the
specific interpretation for their formation.  The presence of a ring
requires either 1) a brief ejection episode as a blue supergiant, B[e]
star, or LBV, where $\dot{M}$ increased drastically for a short time,
or 2) the initial formation of a thin disk over a long timescale,
which is then swept up into a ring following a sharp increase in the
speed of the stellar wind.  Both these mechanisms would require that
the ejection of rings is a consequence of late stages of stellar
evolution, rather than normal main-sequence evolution.

In the context of linking spatially-resolved rings to B[e]
supergiants, RY Scuti (Smith et al.\ 2002b) is perhaps one of the most
interesting objects available for detailed study.  It has a toroidal
nebula, it shares many spectroscopic properties in common with the
B[e] supergiants (Men'shchikov \& Miroshnichenko 2005; Allen \& Swings
1976; Smith et al.\ 1999), and it is an eclipsing binary in a state of
overcontact that is known to be shedding mass and angular momentum.
Its rings have a radius of 1000--2000 AU --- in between the expected
radii of emitting circumstellar material around B[e] stars (a few to
several hundred AU) and most of the spatially-resolved rings discussed
here at radii of a few $\times$10$^4$ AU.  The dust mass and IR excess
of RY Scuti are similar to B[e] supergiants, although the grains are
at a lower temperature around RY Scuti because they are farther from
the star.  The $\sim$40 km s$^{-1}$ expansion speed of its rings
(Smith et al.\ 2002) is typical of the narrow-line regions around B[e]
stars (Zickgraf et al.\ 1986), but faster than the speeds of the more
extended rings discussed in this paper, which have presumably
decelerated.  Thus, RY Scuti may represent a very short-lived
``missing link'' between the B[e] supergiants and the early B
supergiants surrounded by rings.

\subsection{The Role of Binarity}

The sample of objects in Table 3 may yield interesting clues to the
role of close binary interactions in the formation mechanism of
circumstellar rings.  Six of the nine objects are known to be or
suspected to be close interacting binaries.  RY Scuti, SuWt2, and
He2-147 are definite eclipsing binaries, while the central star of
WeBo1 is a Barium star that probably resulted from mass transfer in a
Roche lobe overflow (RLOF) phase (Bond et al.\ 2003).  The remaining
two suspected binaries are somewhat unusual cases: SN1987A is thought
to have undergone a binary merger event when the ring was ejected, as
noted in \S 1 (although there may be problems with the merger
interpretation; Smith 2007; Smith \& Townsend 2007).  The invisible
central star in the SBW2 ring probably requires that a central binary
system was disrupted and that the stars were ejected from the system,
as discussed earlier in \S 3.2.  Taken together, this collection of
objects strongly suggests that mass loss through the outer Lagrangian
point in a common-envelope/RLOF phase of close binary evolution is a
preferred avenue for the forging of rings, at least for
intermediate-mass systems.  There are good reasons why more luminous
massive stars can produce rings without close binary influence (Smith
\& Townsend 2007).  RY Scuti is a key object in this regard, as it is
currently caught in this brief but important RLOF phase and it is the
only confirmed binary among the supergiants with rings.  In a scenario
where a massive star has a companion close enough and massive enough
to shape its ejecta, it would be difficult for that companion to
escape detection.

Forging of such rings through close binary interaction may be relevant
for the formation of bipolar planetary nebulae if a fast wind expands
into such an environment with a strong equatorial density enhancement.
Some of the most tightly-pinched waists among bipolar planetary
nebulae are seen in so-called symbiotic PNe, and several of these are
strongly suspected to be binaries (e.g., M 2-9 [Doyle et al.\ 2000];
Mz~3 [Smith 2003]; Hb12 [Hsia et al.\ 2006]).  However, several of the
objects we discuss here are equatorial rings that are suspected to be
present {\it before} the planetary nebula phase, and are only seen in
rare circumstances as discussed next.

\subsection{A Secret That Only Fire Can Tell: Are Equatorial Rings Commonplace?}

Two-thirds of the objects in Table 3 (the two new objects discovered
here, plus SN1987A, Sher 25, HD168625, and WeBo1) are seen projected
in or near giant H~{\sc ii} regions --- environments that may provide
a copious external source of UV photons (admittedly, the nebula of
SN1987A was flash ionized by the supernova event itself).  RY Scuti
has a sufficiently hot and luminous star to provide its own UV
illumination. SBW2 in Carina has no bright central star, but we have
shown earlier that the hot stars that power the Carina Nebula can
supply sufficient ionizing photons.  SuWt2, with two A-type stars, may
have a similar problem, with a nearby early B star providing a similar
solution.  WeBo1 is near W4 and the X-ray binary LS I +61\arcdeg303.
This raises the question of whether such rings around post-MS stars
might be fairly common, but go unseen where there is a lack of
favorable external illumination.  Such precious rings may be the long
sought-after doughnuts (e.g., Frank et al.\ 1998) needed for the
interacting winds scenario of bipolar planetary nebula formation.
Even if external ionization is not critical in making these rings
visible, the fact that most of them are seen in close proximity to
H~{\sc ii} regions certainly increases their chances for serendipitous
discovery.

This suggests that there may be many more rings of similar size and
morphology surrounding post-MS intermediate-mass stars throughout the
Galaxy, but they go undetected either because they are not illuminated
or because most of the Galactic plane has not been surveyed with deep
narrowband imaging as intensively as star-forming regions have been.
Excluding the more massive stars in our sample, which may supply their
own ionization and are probably found preferentially near H~{\sc ii}
regions anyway, we can make an order of magnitude extrapolation.
H~{\sc ii} regions occupy only 0.1--1\% of the volume of the galactic
disk, so there may be several hundred to a few thousand more dark
rings in the Galaxy, comparable to the known number of planetary
nebulae.  This estimate is probably conservative, since not all
galactic H~{\sc ii} regions have been surveyed in detail.  Thus, it is
at least plausible that such a population of undetected rings could
exist, and that these could play a role in binding the waists of
bipolar planetary nebulae.  Thus, such rings may partly answer the
question ``where's the doughnut'' posed by Frank et al.\ (1998).

\section{SUMMARY}

\subsection{SBW1 in Context}

The discovery of the ring in SBW1 is the most interesting result of
this study, because it adds to the small number of such rings known
around blue supergiants that are likely supernova progenitors.  In
particular, the radius and morphology of the ring in SBW1 make it
nearly identical to the equatorial ring around SN1987A, and the
luminosity and spectral type of the central star are almost identical
to Sk$-$69\arcdeg202.  The other two B supergiants with similar ring
nebulae, Sher 25 and HD~168625, are both significantly more luminous
than Sk$-$69\arcdeg202.  The low N abundance in SBW1, however, is
surprising because it means that the star has not yet been a RSG, and
so hot supergiants must be able to eject rings like this without a
slow/fast interacting wind scenario.  One intriguing possibility is
that these rings correspond to older phases of ejected material that
would have initially been seen closer to the star as a B[e]
supergiant. If IR excess emission were detected in SBW1, it would add
weight to this idea since B[e] supergiants have dusty disks or tori.
The same may be true for Sher 25, HD168625, and the progenitor of
SN1987A, while the much smaller radius rings around RY Scuti may
suggest that it is in an intermediate phase.  SBW1 should be monitored
photometrically and spectroscopically to determine if it is a close
binary system.

\subsection{SBW2 in Context}

In all respects, except for the troublesome fact that its central star
is missing, SBW2 is a carbon copy of SuWt2 and WeBo1.  These rings are
associated with intermediate-mass (2-8 M$_{\odot}$) close binary
stars.  Their rings appear thicker and more diffuse, with inner edges
that are sharper than their outer limits, as compared to the more
elegant equatorial rings around the blue supergiant stars.  Their
radii around 0.2~pc are oddly similar to those surrounding the more
massive stars.  They all seem to be nitrogen rich, with [N~{\sc ii}]
$\lambda$6583 much brighter than H$\alpha$.  All seem to lack a
suitable source of ionizing photons, unless a hot white dwarf has gone
undetected in each system.  We have suggested that the lack of any
bright central star in SBW2 may indicate that a binary system has been
distrupted, which would be quite extraordinary, while the remaining
rings in the intermediate-mass class all seem to be close binaries.
Thus, in contrast to the massive stars with rings, at low/intermediate
mass it seems more likely that close binary evolution rather than
rapid rotation of single stars governs the forging of these rings.  If
they are detected only in these few cases because of selection effects
or favorable ionization, such rings may be numerous in the Galaxy.  As
such, they could provide the pre-existing equatorial density
enhancements needed to form bipolar nebulae through the interacting
winds scenario.

\acknowledgements \scriptsize

We thank Steve Lawrence for kindly obtaining the blue spectrum of SBW1
for us, and Nolan Walborn for advice on its spectral type.  We also
thank Jon Morse for assistance on the Magellan observing run and Bo
Reipurth for assistance with the Keck run.  Partial support for N.S.\
was provided by NASA through grant HF-01166.01A from the Space
Telescope Science Institute, which is operated by the Association of
Universities for Research in Astronomy, Inc., under NASA contract
NAS~5-26555.  Additional support was provided by NSF grant AST
98-19820 and NASA grants NCC2-1052 and NAG-12279 to the University of
Colorado.

\begin{deluxetable}{lcccc}\tabletypesize{\scriptsize}
\tablecaption{Two New Ring Nebulae in Carina}
\tablewidth{0pt}
\tablehead{
 \colhead{Name} &\colhead{$\alpha$(2000)} &\colhead{$\delta$(2000)} 
         &\colhead{Apparent Size} &\colhead{Comments } }
\startdata

SBW1	&10$^{\rm h}$40$^{\rm m}$19$\fs$4	&-59$\arcdeg$49$\arcmin$09$\farcs$7	
	&11$\farcs$1$\times$7$\farcs$1		&$V$=12.7, $R$=12.14	\\

SBW2	&10$^{\rm h}$36$^{\rm m}$09$\fs$2	&-59$\arcdeg$09$\arcmin$19$\farcs$2	
	&37\arcsec$\times$23$\arcsec$		&($R$=18.5, 20.5, 20.7)	\\

\enddata \tablecomments{For SBW1 these coordinates correspond to the
bright central star, while the position for SBW2 corresponds to the
approximate center of the ring, since no obvious bright star can be
associated with it.  $V$ and $R$ magnitudes of the stars are from the
GSC 2.2 catalog.}
\end{deluxetable}

\begin{deluxetable}{lccccc}\tabletypesize{\scriptsize}
\tablecaption{Dereddened Line Intensities\tablenotemark{a}}
\tablewidth{0pt}
\tablehead{
 \colhead{$\lambda$(\AA)} &\colhead{I.D.} &\colhead{SBW1} 
         &\colhead{SBW2} &\colhead{SuWt2} &\colhead{WeBo1} }
\startdata

5755	&[N~{\sc ii}]	&$<$0.99	&$<$17.1	&11.3	&$\la$2.4	\\
5876	&He~{\sc i}	&\nodata	&\nodata	&...	&5.7	\\
6300	&[O~{\sc i}]	&\nodata	&30.4   	&20.7	&14.6	\\
6312	&[S~{\sc iii}]	&\nodata	&\nodata	&10.5	&...	\\
6364	&[O~{\sc i}]	&\nodata	&\nodata   	&\nodata&11.9	\\
6548	&[N~{\sc ii}]	&14.1   	&174    	&201	&62.2	\\
6563	&H$\alpha$	&100    	&100    	&100	&100	\\
6583	&[N~{\sc ii}]	&46.2   	&568    	&620	&201	\\
6717	&[S~{\sc ii}]	&6.82   	&31.1   	&46.3	&4.7	\\
6731	&[S~{\sc ii}]	&6.56   	&26.9   	&34.8	&4.1	\\
\cutinhead{Derived (assumed) properties}
E(B-V)	&\nodata	&(1.0)		&(0.5)		&(0.25)		&(0.57)	\\
n$_e$	&[S~{\sc ii}]	&513 cm$^{-3}$	&280 cm$^{-3}$	&90 cm$^{-3}$	&312 cm$^{-3}$	\\
T$_e$	&[N~{\sc ii}]	&$<$12,350 K	&$<$14,900 K	&11,400 K	&$\la$9,600 K	\\
n(N$^+$)/n(H$^+$) &\nodata &$>$1.6$\times$10$^{-5}$ &$>$1.35$\times$10$^{-4}$
                  &2.5$\times$10$^{-4}$	&$\ga$1.3$\times$10$^{-4}$	\\

\enddata \tablenotetext{a}{Intensities are given relative to
H$\alpha$=100, and have been dereddened using $R$=3.1 and the listed
$E(B-V)$ values.  Uncertainties in fainter lines are typically 10\%,
but more for the faint lines in WeBo1.  Upper limits are given for
[N~{\sc ii}] $\lambda$5755 in SBW1 and SBW2, while the value for WeBo1
is a marginal ($\sim$2$\sigma$ detection), which we regard as an
approximate upper limit.}
\end{deluxetable}

\begin{deluxetable}{lcccccccc}\tabletypesize{\scriptsize}
\tablecaption{Properties of Well-Defined Equatorial Ring Nebulae}
\tablewidth{0pt}
\tablehead{
 \colhead{Name} &\colhead{Radius} &\colhead{Inclination} &\colhead{V$_{\exp}$\tablenotemark{a}}
         &\colhead{Age (yr)} &\colhead{HII region} &\colhead{N rich?} &\colhead{Binary?} &\colhead{Comments } }
\startdata

SBW1	&0.19 pc	&50$\fdg$2	&19 	&9600 	&Carina		&N	&?		&B1.5 Iab 	\\

SBW2	&0.22 pc	&52\arcdeg\	&$<$8 	&$>$2.6(4) &Carina	&Y	&disrupted?	&no star	\\

SN1987A	&0.21 pc	&44\arcdeg\	&10 	&2(4) 	&30 Dor		&Y	&merged?	&B3 I; triple rings	\\

Sher 25	&0.20 pc	&64\arcdeg\	&20 	&9600 	&N3603		&Y	&...		&B1.5 Ia	\\

HD168625&0.09 pc	&61\arcdeg\	&19 	&4500 	&M17		&Y	&...		&LBV; triple rings	\\

WeBo1	&0.24 pc	&69\arcdeg\	&?	&... 	&W4		&Y	&mass transfer	&Barium star	\\

SuWt2	&0.2 pc ($D_{\rm kpc}$) &64\arcdeg\	&?	&... 	&...	&Y	&eclipsing	&A+A binary	\\

He2-147	&0.025 pc	&55\arcdeg\	&50	&480 	&...		&Y	&eclipsing	&symbiotic mira	\\

RY Sct	&0.005 pc	&75\arcdeg\	&43	&110	&...		&Y	&eclipsing	&WR+OB; double rings	\\

\enddata
\tablenotetext{a}{This is the {\it radial} expansion velocity in km s$^{-1}$
corrected for the inclination of the system.}

\tablerefs{
SN1987A: Plait et al.\ (1995); Crotts \& Heathcote (2000) --- 
Sher~25: Brandner et al.\ (1997) --- 
HD168625: see Smith (2007) and references therein --- 
WeBo1: Bond et al.\ (2003) ---
SuWt2: Schuster \& West (1976) and this work --- 
He~2-147: Corradi et al.\ (1999, 2000) --- 
RY~Sct: Smith et al.\ (1999, 2001, 2002$b$).
}
\end{deluxetable}


\begin{figure}
\epsscale{0.55}
\plotone{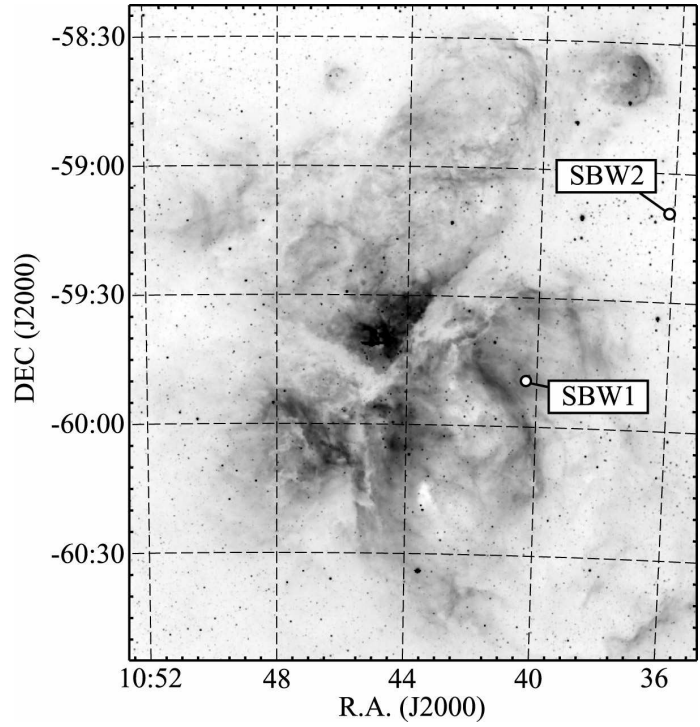}
\caption{Large-scale H$\alpha$ image of the Carina Nebula (from Smith
et al.\ 2000) showing the locations of SBW1 and SBW2 relative to the
rest of the giant H~{\sc ii} region.}
\end{figure}

\begin{figure}
\epsscale{0.9}
\plotone{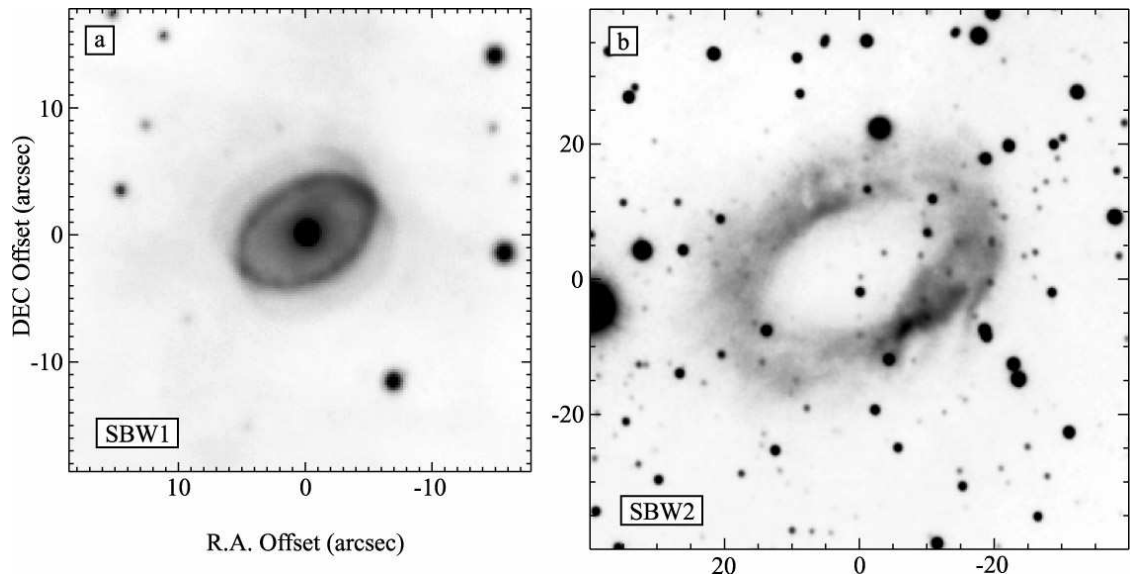}
\caption{Magellan/MagIC H$\alpha$+[N~{\sc ii}] images of SBW1 (a) and
SBW2 (b).  Note that the two images are not reproduced here at exactly
the same size scale.}
\end{figure}

\begin{figure}
\epsscale{0.8}
\plotone{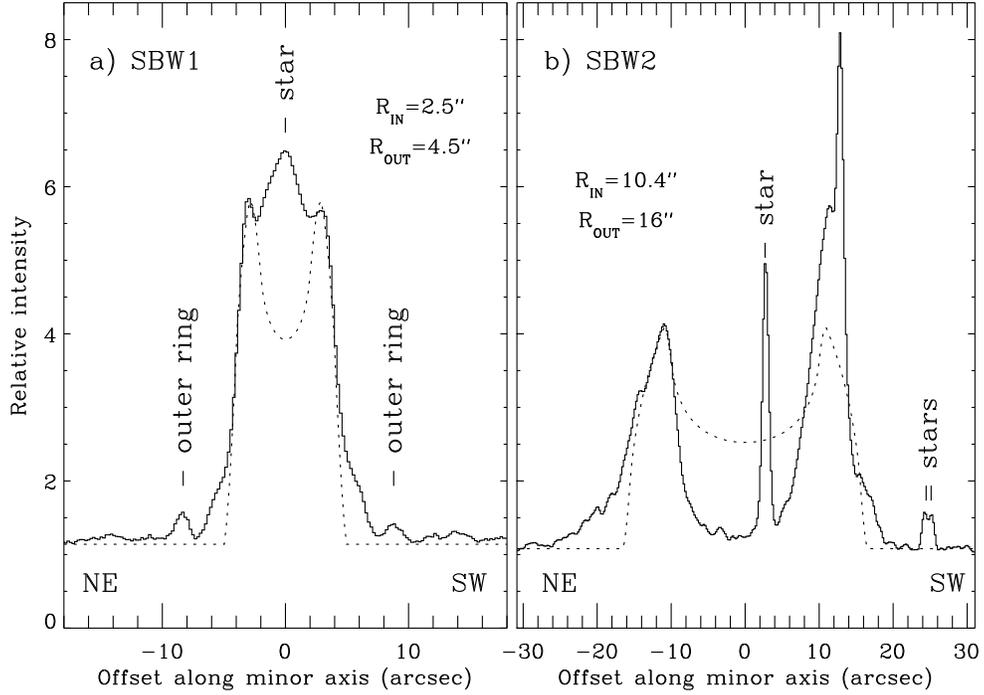}
\caption{Intensity tracings across the images in Figure 2.  (a)
Tracing along the minor axis of SBW1, taken from two 0$\farcs$5-wide
segments on either side of the EMMI slit aperture shown in Figure
4$a$, in order to avoid bright light from the star.  (b) Tracing
through the middle of SBW2, immediately to the northwest of the EMMI
slit shown in Figure 4$b$.  In each panel, the dashed line is a model
for the expected emission from a cross section through a
limb-brightened thin shell with the inner and outer radii listed in
each panel.  In Panel B, especially, it is clear that the observed
emission is not consistent with the limb-brightened shell.}
\end{figure}

\begin{figure}
\epsscale{0.9}
\plotone{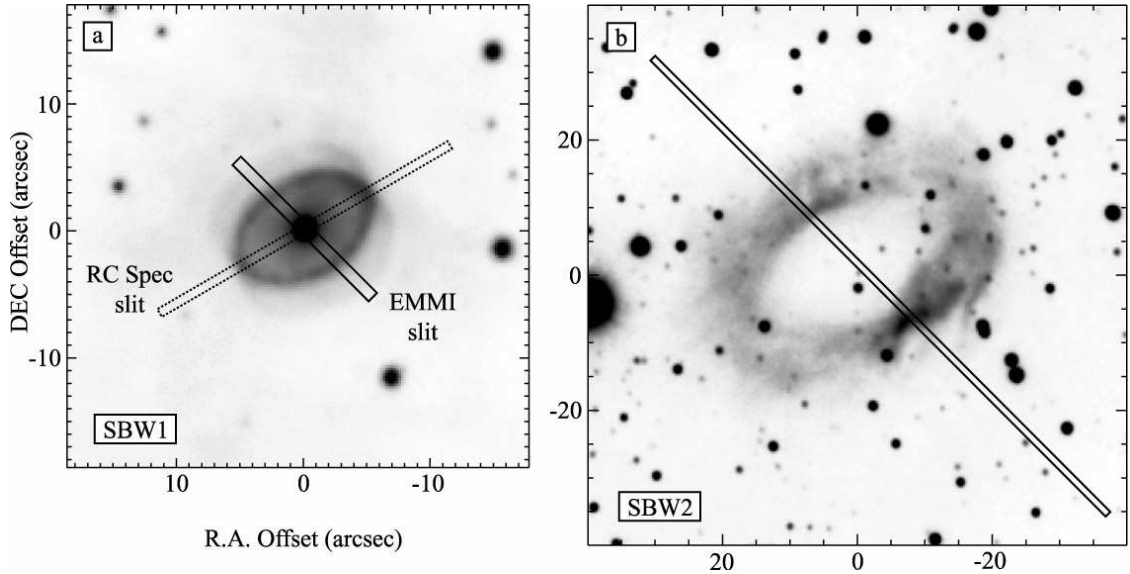}
\caption{Same as Figure 2, but showing the EMMI and RC Spec slit
positions.}
\end{figure}

\begin{figure}
\epsscale{0.7}
\plotone{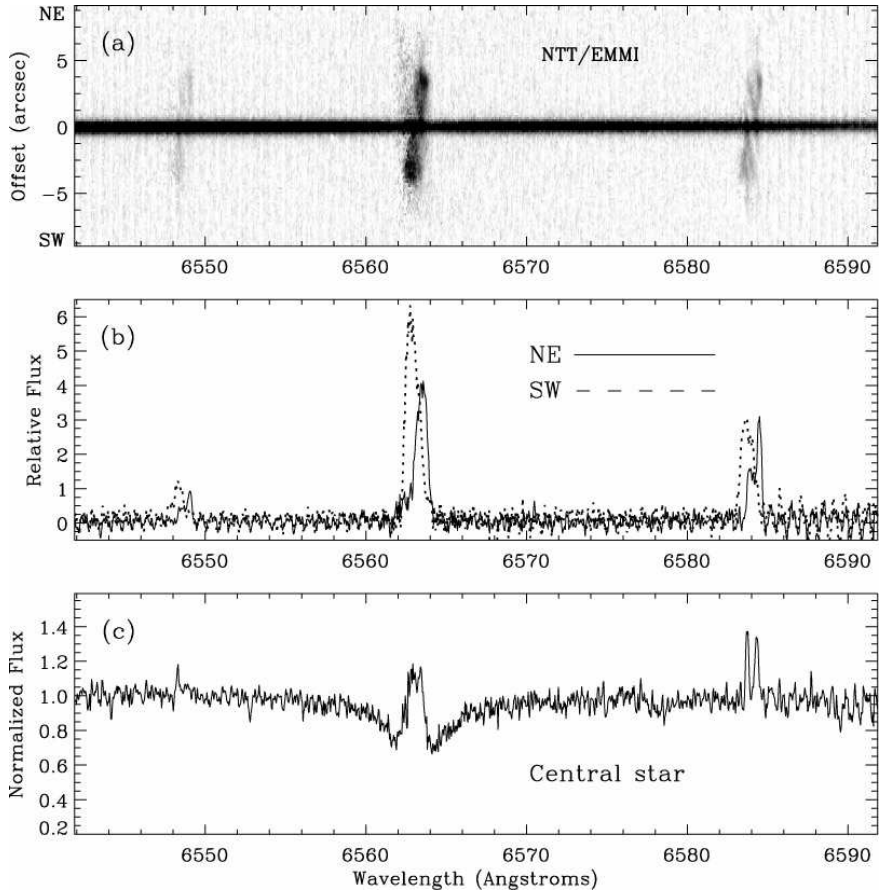}
\caption{(a) Background-subtracted long-slit spectrum of SBW1 in the
region near H$\alpha$, with the 15\arcsec\ slit aperture of EMMI
passing through the central star, oriented at P.A.=45\arcdeg\ (NE is
up and SW is down).  (b) Intensity tracings of the nebular emission in
Panel A, offset to the NE (solid) and SW (dashed). (c) Extracted
spectrum of the central star in the same wavelength range; background
H~{\sc ii} region emission has been subtracted, but circumstellar
nebular emission has not.}
\end{figure}

\begin{figure}
\epsscale{0.5}
\plotone{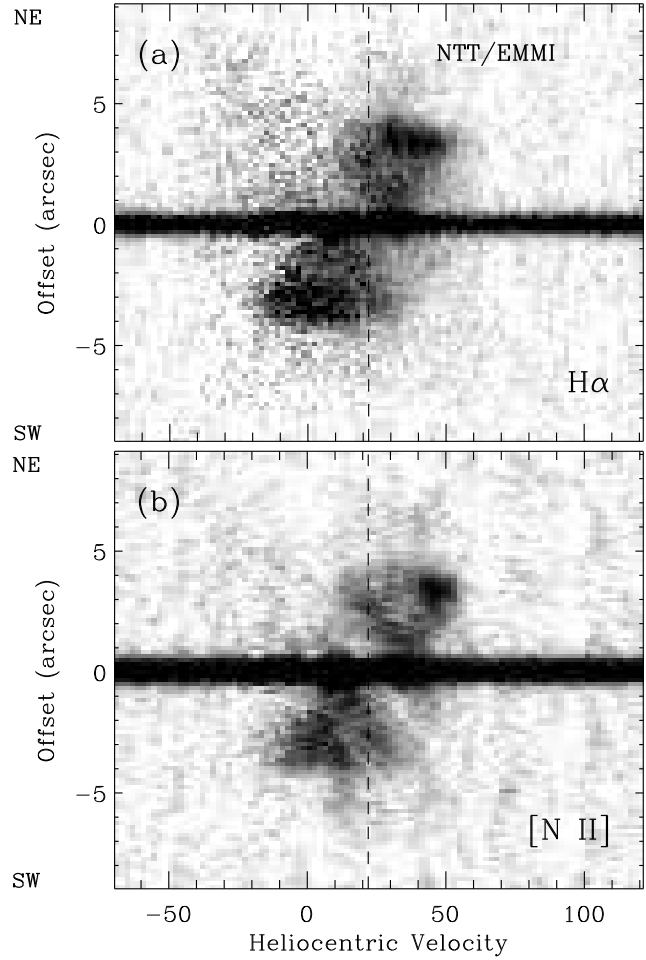}
\caption{Long-slit nebular kinematics of SBW1 in (a) H$\alpha$ and (b)
[N~{\sc ii}] $\lambda$6583.  The velocity scale is heliocentric, and
the dashed line marks the presumed systemic velocity at +22 km
s$^{-1}$, which is offset by +30 km s$^{-1}$ from the average emission
velocity of gas in the Carina Nebula at $-$8.1 km s$^{-1}$ (Smith
2004).  These two panels are sections of the data in Figure 5$a$.}
\end{figure}

\begin{figure}
\epsscale{0.7}
\plotone{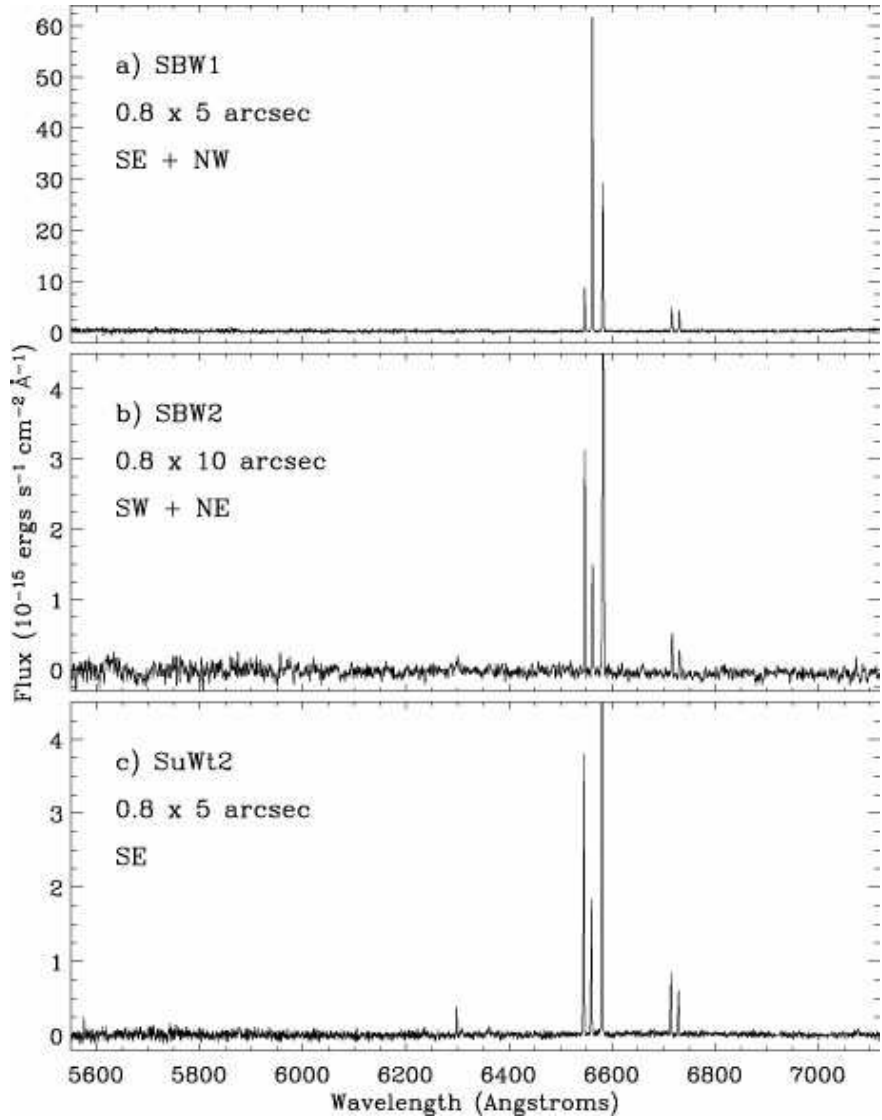}
\caption{Extracted CTIO/4m RC Spec spectra of SBW1 (a), SBW2 (b),
and SuWt2 (c).  Aperture sizes and their approximate locations are
noted in each panel.}
\end{figure}

\begin{figure}
\epsscale{0.6}
\plotone{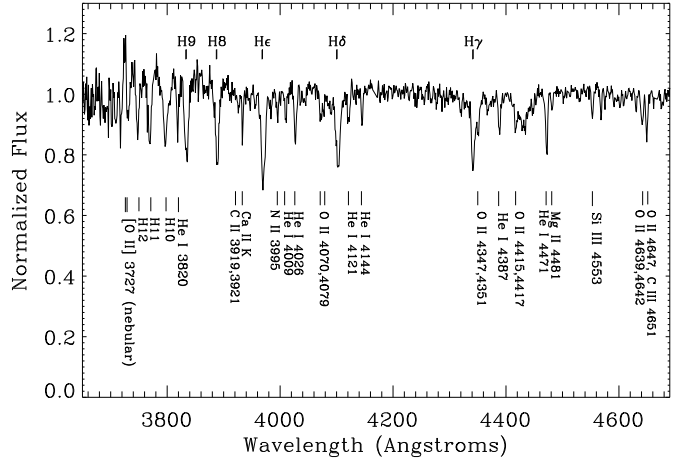}
\caption{Blue spectrum of the central star of SBW1.  The MK spectral
type is approximately B1.5 Iab.}
\end{figure}

\begin{figure}
\epsscale{0.9}
\plotone{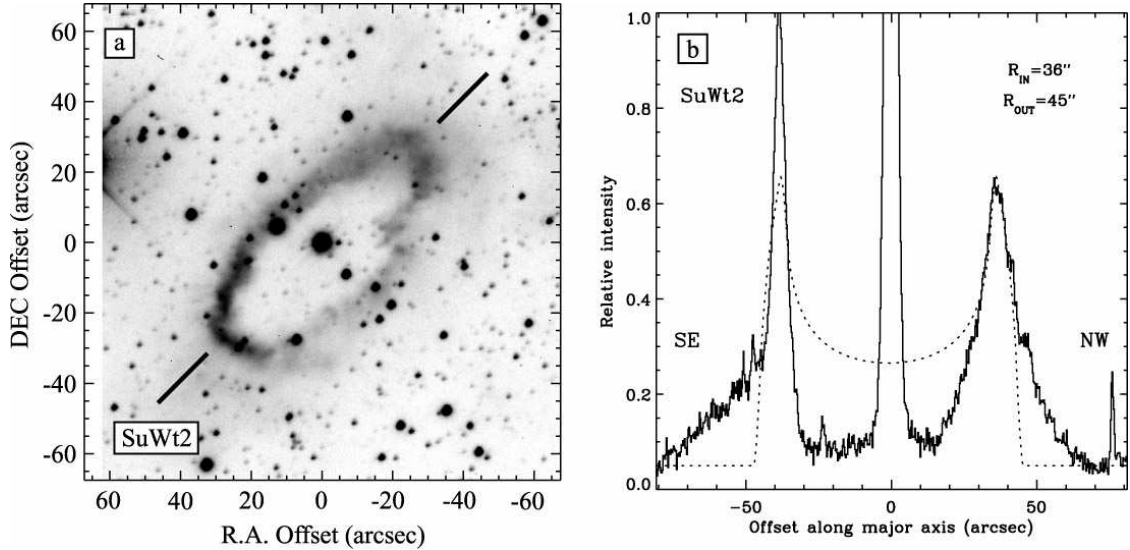}
\caption{(a) H$\alpha$+[N~{\sc ii}] image of SuWt2 obtained with the
MOSAIC2 camera on the CTIO 4m telescope.  (b) Same as in Figure 3, but
for SuWt2.  The intensity tracing cuts across the major axis of the
ring (along P.A.=--45$\arcdeg$).  The diagonal lines in Panel $a$ mark
the orientation of the slit for the RC Spec observations, as well as
the scan direction for Panel $b$.  The central star is offset from the
exact center of the ring by $\sim$1\arcsec\ to the NW.}
\end{figure}

\begin{figure}
\epsscale{0.9}
\plotone{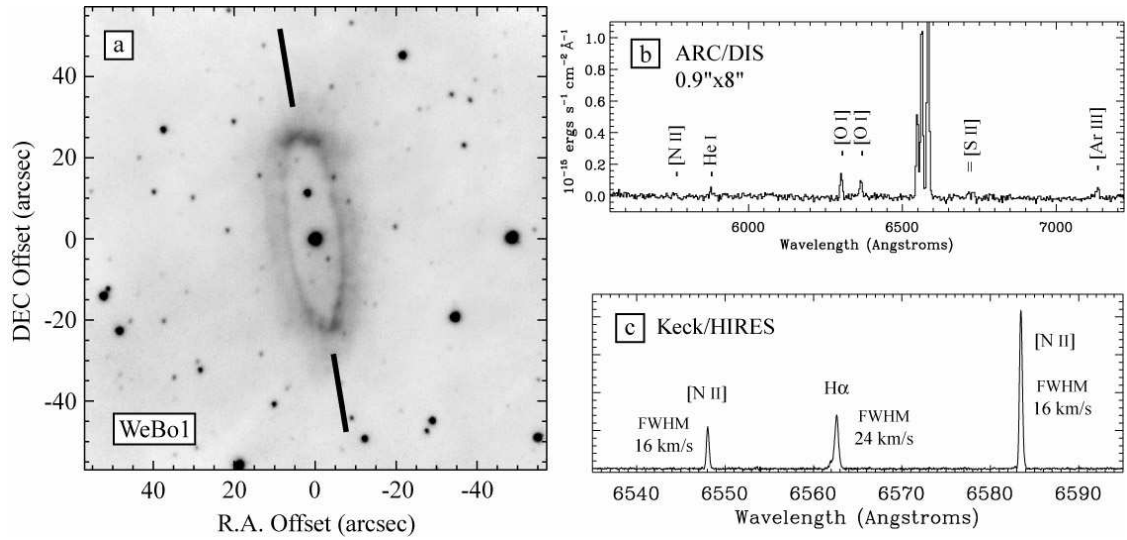}
\caption{(a) H$\alpha$+[N~{\sc ii}] image of WeBo1 obtained with
SPICam on the ARC 3.5m telescope.  (b) A low resolution red spectrum
of WeBo1 obtained with DIS on the ARC telescope.  (c) a higher
resolution spectrum of the H$\alpha$+[N~{\sc ii}] lines obtained with
the HIRES spectrograph on Keck.}
\end{figure}

\end{document}